\documentclass[aps,prc,twocolumn,showpacs,superscriptaddress,footinbib]{revtex4-1}

\usepackage{graphicx}
\usepackage{xcolor}
\usepackage{dcolumn}
\usepackage{bm}
\usepackage{enumerate}
\usepackage{amsmath}

\usepackage[normalem]{ulem}  

\begin{document}


\title{Typical new spherical-like $\gamma$-soft spectra in $^{104,106,108}$Pd}

\author{Tao Wang}
\email{suiyueqiaoqiao@163.com}
\affiliation{College of Physics, Tonghua Normal University, Tonghua 134000, People's Republic of China}

\date{\today}

\begin{abstract}
To solve the Cd puzzle (spherical nucleus puzzle), I have proposed a new spherical-like $\gamma$-soft nucleus. Since shape coexistence often occurs in such nuclei, explicit spherical-like $\gamma$-soft spectra are not easily identified. In this paper, I finally find  the direct evidence for the existence of the new spherical-like $\gamma$-soft nucleus. $^{106}$Pd is in fact a typical spherical-like $\gamma$-soft nucleus. The low-lying parts, up to the $10_{1}^{+}$ state, under 4000 keV, of the spherical-like $\gamma$-soft spectra are verified. The B(E2) values and the quadrupole moments of the low-lying levels are also studied. By comparison, the new spherical-like $\gamma$-soft mode is confirmed. Moreover $^{104,108}$Pd are also discussed. These results completely disprove the possibility of the phonon excitations of the spherical nucleus in the Cd-Pd nuclei region.
\end{abstract}

\maketitle

\section{Introduction}

The spherical nucleus, as well as its surface vibrational excitations, as one of the main paradigms of the collective excitations in nuclear structure, has been believed for decades. However, this traditional idea has been questioned recently (spherical nucleus puzzle) \cite{Garrett10,Heyde11,Garrett16,Heyde16,Garrett18}. Based on the experimental data from the Cd nuclei (Cd puzzle) \cite{Garrett08,Garrett12,Batchelder12,Garrett19,Garrett20}, especially on the observation that there is no the $0_{3}^{+}$ state at the three-phonon levels, I have proposed a new spherical-like $\gamma$-soft nucleus \cite{Wang22}, which is a $\gamma$-soft rotational mode with spherical-like spectra for the low-lying part. The experimental analysis of the existence of the new spherical-like $\gamma$-soft spectra can be seen in Ref. \cite{Wang25}.

An extension of the interacting boson model (IBM) \cite{Iachello87,Iachello75} with SU(3) higher-order interactions was proposed by me \cite{Wang22} to produce the lowest part of the spherical-like spectra successfully \cite{Wang22,Wang25}. In the IBM, discussions of the collective behaviors are based on the four dynamical symmetry limits: the U(5) symmetry limit (spherical shape), the SU(3) symmetry limit (prolate shape), the O(6) symmetry limit ($\gamma$-soft rotation), and the $\overline{\textrm{SU(3)}}$ symmetry limit (oblate shape) \cite{Jolie01,Casten10,Wang08}. However, in the new model, the SU(3) symmetry is the most important, and dominates all the quadrupole deformations including the oblate shape \cite{Wang22,Wang25}. Historically, SU(3) symmetry has played a very important role in the researches of nuclear structure \cite{Elliott581,Arima69,Hecht69,Draayer83,Deaayer87,Isacker85,Isacker00,zhang14,Bonatsos17,Kota20}. Now, it becomes more important, or even fundamental.

\begin{figure}[tbh]
\includegraphics[scale=0.33]{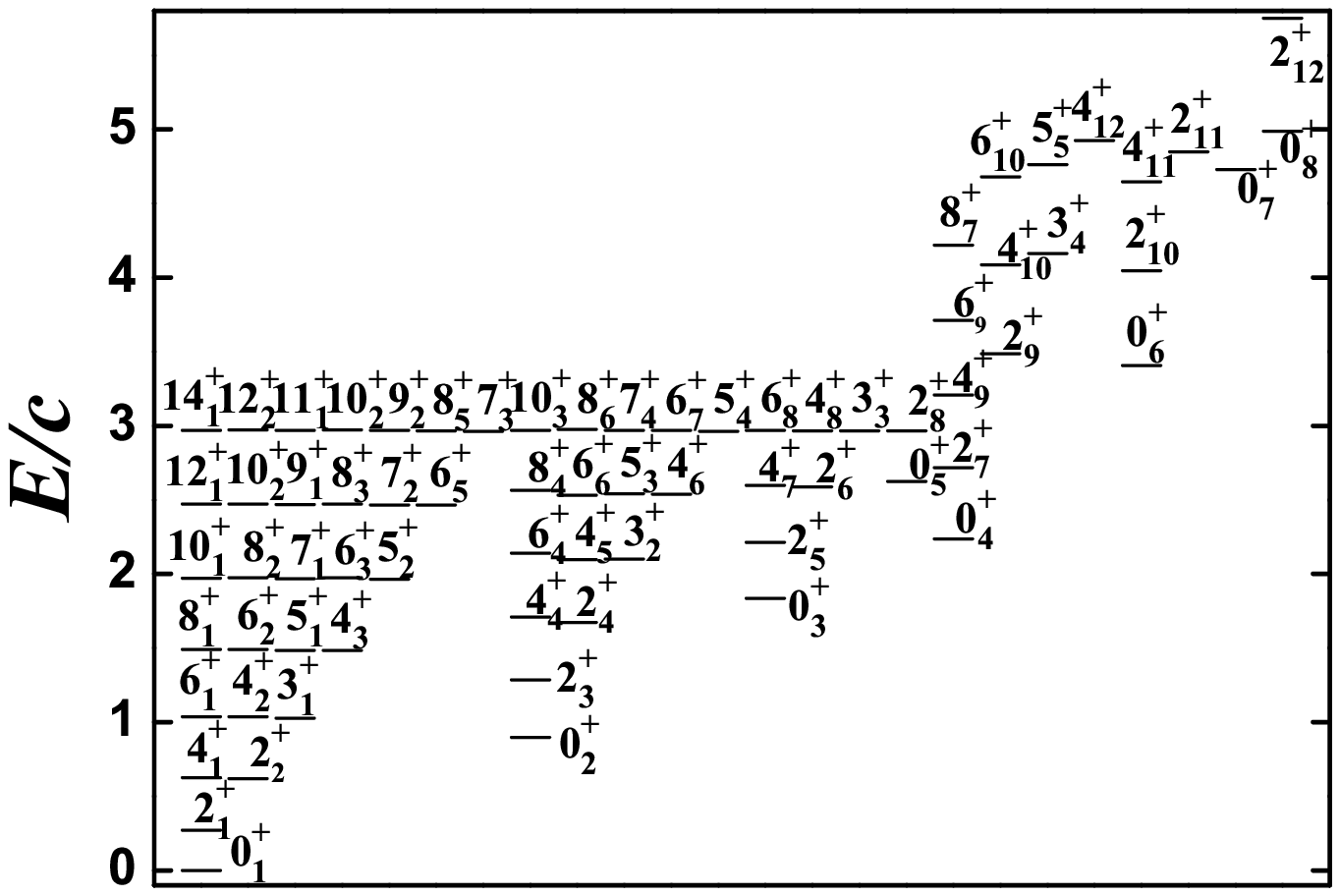}
\caption{New spherical-like $\gamma$-soft spectra for $N=7$, which is the boson number of $^{106}$Pd.}
\end{figure}

The new model can be also used to explain the B(E2) anomaly \cite{Wang20,Zhang22,Wangtao,Zhang24,Pan24,Zhang25,Zhang252,Zhang253,Teng25,Cheng25,Li25}, to explain the prolate-oblate shape asymmetric evolution \cite{Fortunato11,Zhang12,Wang23}, to produce the features of the $\gamma$-soft nucleus $^{196}$Pt at a better level \cite{WangPt}, and to produce the E(5)-like spectra in $^{82}$Kr \cite{Zhou23}. Recently this model can successfully produce the boson number odd-even effect in $^{196-204}$Hg \cite{WangHg}, which is a unique prediction of the new model \cite{Zhang12}. Otsuka \emph{et al.} argued that nuclei previously considered prolate shape should be rigid triaxial \cite{Otsuka19,Otsuka21,Otsuka24}. Recently, this result has been confirmed in $^{166}$Er with the new model \cite{ZhouEr}. The shape phase transition from the new $\gamma$-soft phase to the prolate shape was also found \cite{Zhao251}. These results prove that the new model can give more accurate descriptions of the collective behaviors in nuclei than previous theories, and can provide a unified picture for understanding the collective excitations in nuclear structure.

\begin{figure}[tbh]
\includegraphics[scale=0.33]{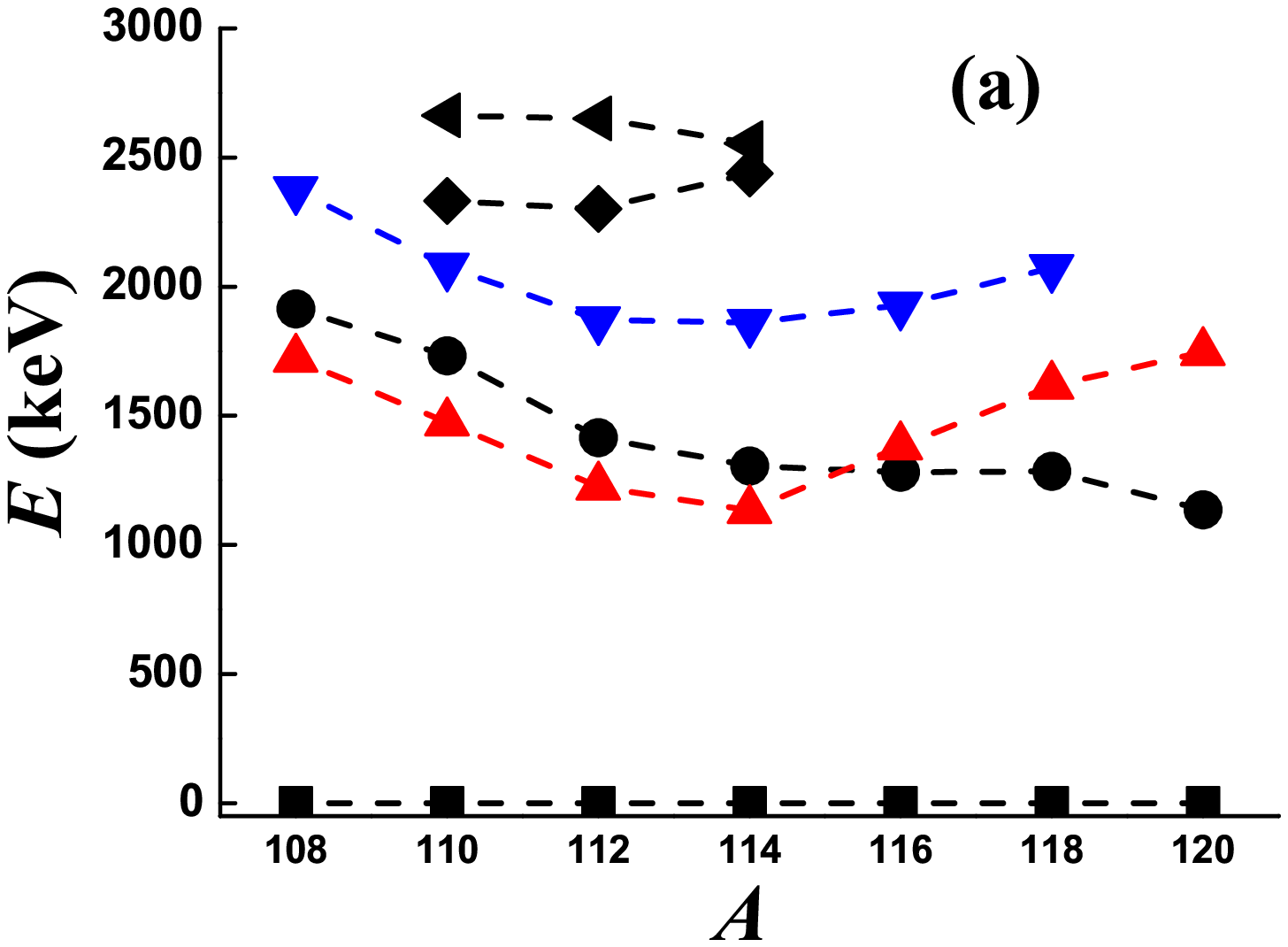}
\includegraphics[scale=0.33]{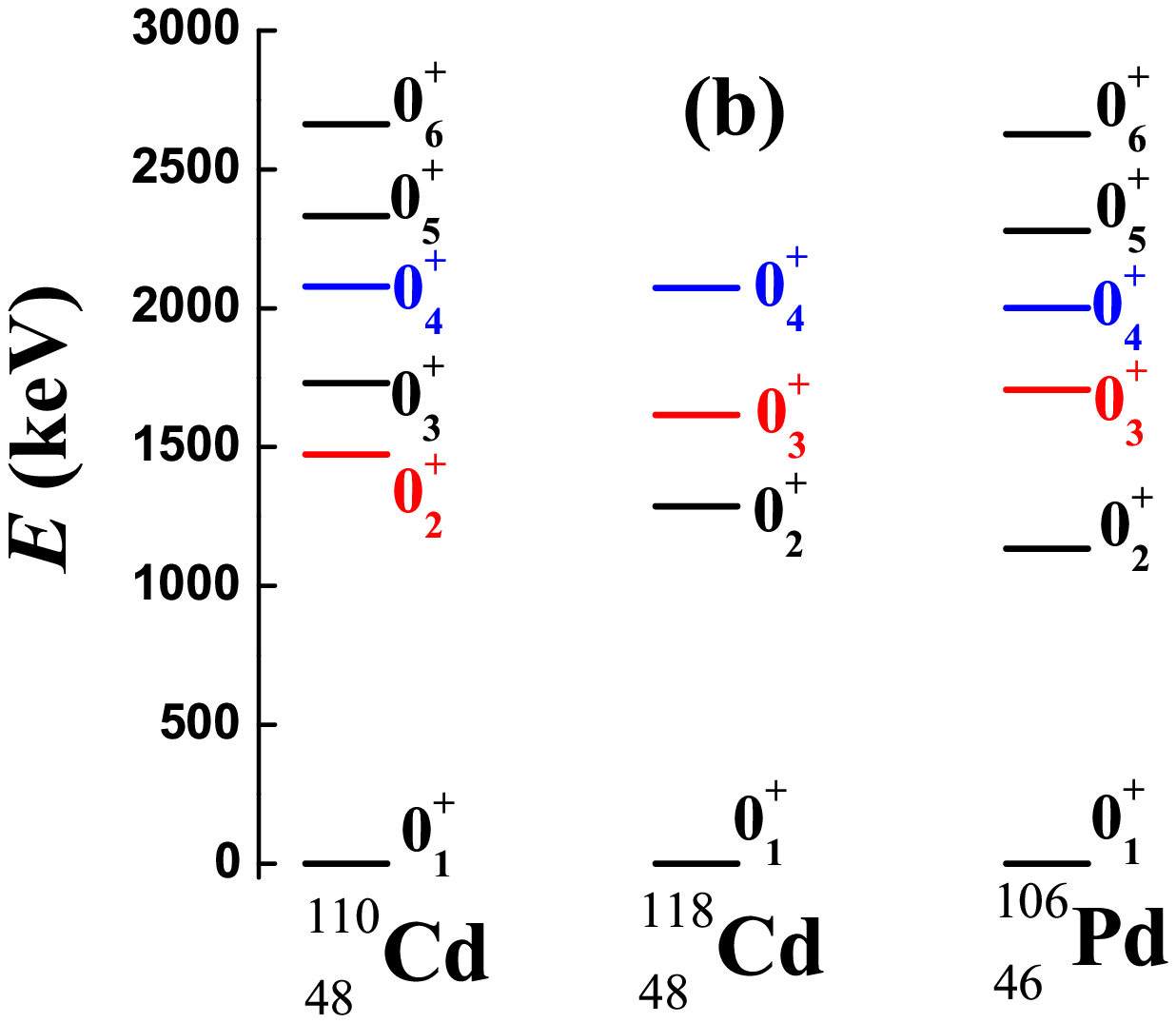}
\caption{(a) The lowest six $0^{+}$ states in Cd nuclei from 108 to 120; (b) The lowest six $0^{+}$ states in $^{110}$Cd, $^{118}$Cd and $^{106}$Pd.}
\end{figure}

Fig. 1 depicts the new spherical-like $\gamma$-soft spectra for $N=7$, which is the boson number of $^{106}$Pd. In Ref. \cite{WangPt}, the spectra for $N=6$ is also shown. For the low-lying part up to the $12_{1}^{+}$ state, the new spherical-like $\gamma$-soft spectra are nearly the same, and resemble the phonon excitations of the spherical nucleus. The characteristics of the spectra are: (1) similar to the two-phonon excitations in the spherical nucleus, the $4_{1}^{+}$, $2_{2}^{+}$, $0_{2}^{+}$ triple states really exist (the first group states); (2) there is no the $0_{3}^{+}$ state near the $6_{1}^{+}$, $4_{2}^{+}$, $3_{1}^{+}$ and $2_{3}^{+}$ states (the second group states), which was discussed in detail in \cite{Wang25}; (3) the $0_{3}^{+}$ state is near the $8_{1}^{+}$, $6_{2}^{+}$, $5_{1}^{+}$, $4_{3}^{+}$, $4_{4}^{+}$ and $2_{4}^{+}$ states (the third group states) whose energy is nearly twice the one of the $0_{2}^{+}$ state (two-times relationship). To the best of my knowledge, this new $\gamma$-soft spectra have never been found in previous models in nuclear structure.

The (1) and (2) features can be found in $^{120}$Cd \cite{Batchelder12} and $^{112}$Cd \cite{Garrett19,Garrett20} experimentally, and can be well described by the new model \cite{Wang22,Wang25}. Especially for the normal states of $^{108-120}$Cd, a systematic fitting has been performed with a single Hamiltonian, and the anomalous evolution trend of the quadrupole moments of the $2_{1}^{+}$ states was first illustrated by the theory \cite{Wang25}. To the best of my knowledge, this phenomenon has not been mentioned in previous studies. However the higher levels $8_{1}^{+}$, $6_{2}^{+}$, $5_{1}^{+}$, $4_{3}^{+}$, $4_{4}^{+}$, $2_{4}^{+}$ and $0_{3}^{+}$ (the third group states) and $10_{1}^{+}$, $8_{1}^{+}$, $7_{1}^{+}$, $6_{3}^{+}$, $5_{2}^{+}$, $6_{4}^{+}$, $4_{5}^{+}$, $3_{2}^{+}$, $2_{5}^{+}$ and $0_{4}^{+}$ (the fourth group states) are not confirmed so far, and the feature of the energy positions of the $0^{+}$ states in the new spherical-like $\gamma$-soft spectra is also not found in actual nuclei.

\begin{figure}[tbh]
\includegraphics[scale=0.33]{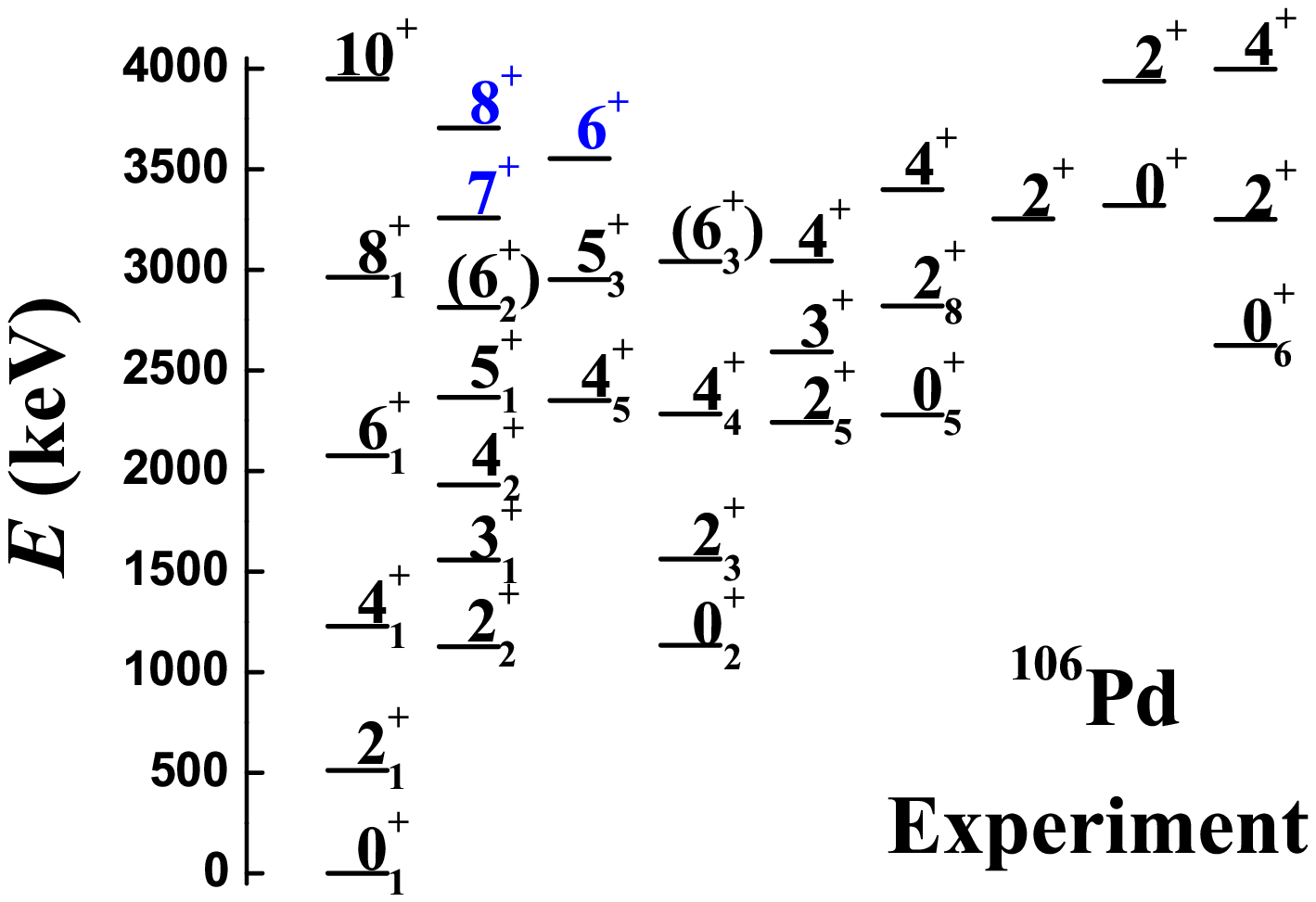}
\includegraphics[scale=0.33]{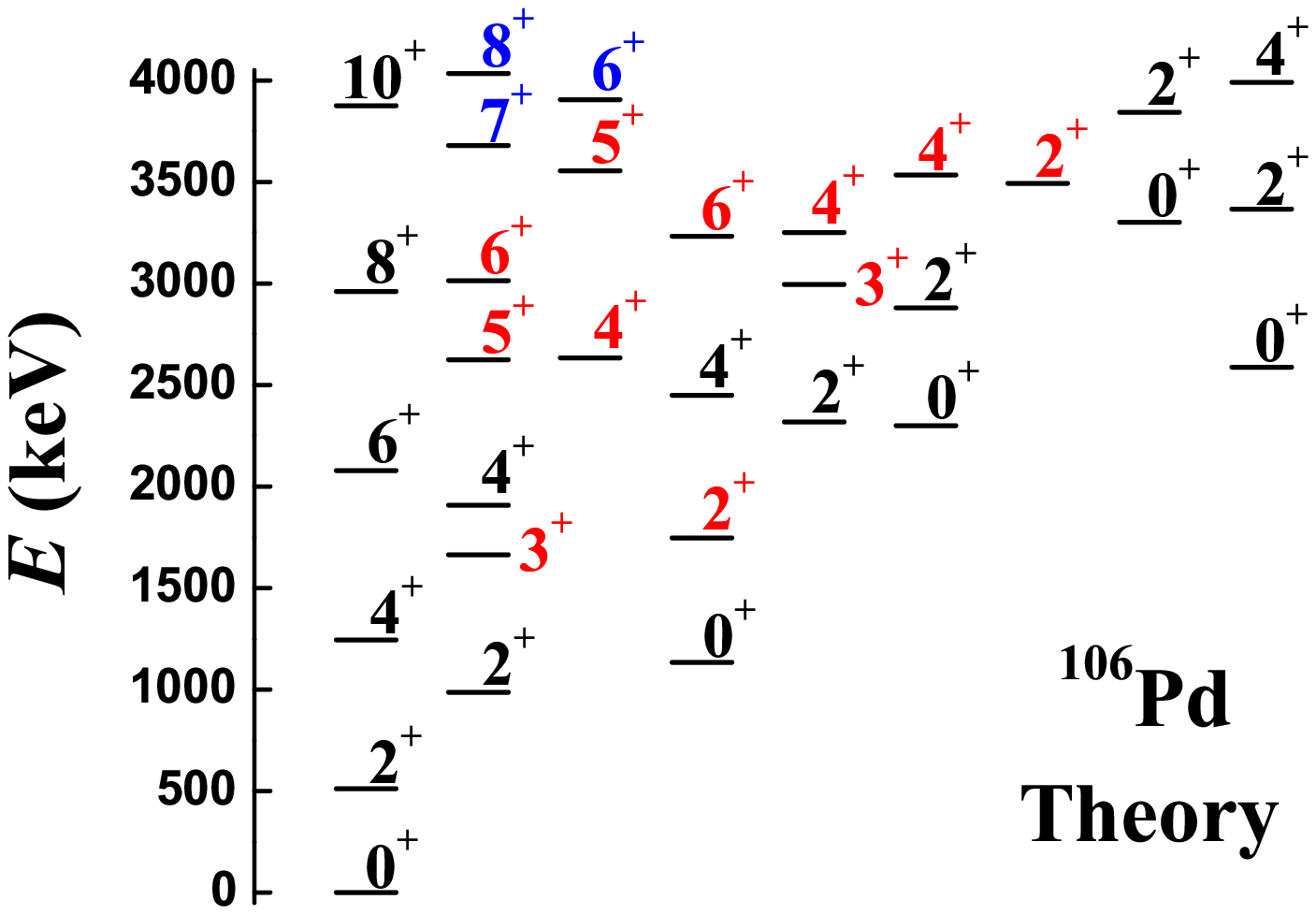}
\caption{The experimental and theoretical results of the low-lying levels of the normal states in $^{106}$Pd.}
\end{figure}

Observing these features experimentally is extremely critical to confirm the existence of the new spherical-like $\gamma$-soft nucleus. The lowest part (up to the $6_{1}^{+}$, $4_{2}^{+}$, $3_{1}^{+}$ and $2_{3}^{+}$ states) only gives the partial characteristics of the new pattern, which is not comprehensively enough, and other possibilities may exist. A direct comparison between experiments and theories is still needed and extremely critical. This can convince the researchers in the field of nuclear structure that there is indeed a new collective excitation, not far from the magic nuclei. In this paper, I show that $^{106}$Pd is really the new spherical-like $\gamma$-soft nucleus, in which the third and fourth group states can be confirmed. The two adjacent nuclei $^{104,108}$Pd are also discussed.

\section{Discussion}

\begin{table}[tbh]
\caption{\label{table:expee}  Absolute $B(E2)$ values in W.u. for $E2$ transitions from the low-lying states in $^{106}$Pd. The theory has effective charge $e=2.027$ (W.u.)$^{1/2}$. $^{a}$From Ref. \cite{ensdf}, $^{b}$From Ref. \cite{Svensson95}, $^{c}$From Ref. \cite{Pd17}, $^{d}$From Ref. \cite{Giannatiempo98,Giannatiempo18}.}
\setlength{\tabcolsep}{0.8mm}{
\begin{tabular}{ccccccc}
\hline
\hline
$L_{i}  ~~L_{f}    $ &Exp. 1$^{a}$                     \ &Exp. 2$^{b}$                \ &Exp. 3$^{c}$                      \ &Theo.   \ & IBM-2$^{d}$   \\
 \hline
$2_1^+  ~~ 0_1^+   $ & 44.3(15)                 \ & 42(4)       \ &  44.3(15)                        \ & 44.3     \ &  40.0   \\
$2_2^+  ~~ 2_1^+   $ & 44(4)                   \ & 39(4)                 \ & 43.7$_{-50}^{+58}$            \ & 49.9       \ &  46.0  \\
$~~~~   ~~  0_1^+  $ & 1.17(10)                \ & 0.87$_{-9}^{+10}$               \ & 1.18$_{-13}^{+15}$            \ & 0.31       \ &  0.52     \\
$0_2^+  ~~ 2_1^+   $ & 35(8)                  \ & 43$_{-9}^{+6}$          \ &   35(8)                      \ & 35.0       \ &  44.7    \\
$~~~~   ~~ 2_2^+   $ &                          \ & 19$_{-3}^{+7}$         \ &                         \ & \textbf{84.6}       \ &  \textbf{27.6}     \\
$4_1^+  ~~  2_1^+  $ & 76(11)                  \ & 71(7)                   \ &  76(11)                       \ & 65.4       \ &  64.5   \\
$~~~~   ~~ 2_2^+   $ &                          \ & 0.7$_{-3}^{+72}$        \ &                         \ & 6.2       \ &   0.98    \\
$3_1^+  ~~  2_1^+  $ &                          \ &                        \ & 0.444$_{-50}^{+57}$          \ & 0.50       \ & 0.92    \\
$~~~~   ~~ 2_2^+   $ &                          \ &                    \ & 16.2$_{-26}^{+30}$           \ & 50.2       \ &  30.1     \\
$~~~~   ~~ 4_1^+   $ &                          \ &                    \ & 6.0$_{-60}^{+11}$           \ & 7.21       \ &  9.38     \\
$2_3^+  ~~ 0_1^+   $ &                          \ & 0.14(2)           \ & 0.147$_{-20}^{+25}$           \ & 0.03      \ &  0.23       \\
$~~~~   ~~ 2_1^+   $ &                          \ & 0.52$_{-7}^{+10}$  \ & 0.52$_{-10}^{+13}$           \ & 0.25       \ & 0.13      \\
$~~~~   ~~  2_2^+  $ &                          \ & 10.2$_{-15}^{+22}$    \ & 10$_{-10}^{+2}$           \ & 5.57          \ &  2.80         \\
$~~~~   ~~  0_2^+  $ &                          \ & 39(4)              \ & 38$_{-11}^{+13}$          \ & \textbf{37.8}       \ & \textbf{12.3 }       \\
$~~~~   ~~  4_1^+  $ &                          \ & 5.3$_{-14}^{+25}$      \ & 10.6$_{-42}^{+51}$          \ & \textbf{22.7}       \ &   \textbf{4.70}      \\
$4_2^+  ~~ 2_1^+   $ &                          \ & 0.007$_{-3}^{+6}$           \ &                        \ & 2.87   \ &  0.001  \\
$~~~~   ~~ 2_2^+   $ & 35(6)                   \ & 35$_{-4}^{+5}$           \ & 34.5$_{-54}^{+71}$            \ & 28.0       \ &  35.2        \\
$~~~~   ~~  4_1^+  $ & 21(3)                   \ & 23$_{-2}^{+3}$           \ & 22.1$_{-39}^{+51}$           \ & 17.4      \ &  25.9      \\
$~~~~   ~~  3_1^+  $ &                          \ &                     \ & 33$_{-33}^{+11}$            \ & \textbf{31.8}      \ &   \textbf{4.61}     \\
$6_1^+  ~~ 4_1^+   $ & 88(9)                   \ & 89$_{-13}^{+10}$         \ & 88.3$_{-79}^{+97}$            \ & 80.2       \ &  75.7      \\
$2_5^+  ~~ 0_1^+   $ &                          \ &                     \ & 0.062$_{-21}^{+25}$           \ & 0.00     \ &   0.00      \\
$~~~~   ~~ 2_1^+   $ &                          \ &                     \ & 0.167$_{-69}^{+83}$          \ & 0.01       \ & 0.01      \\
$~~~~   ~~  2_2^+  $ &                          \ &                     \ & 3.2$_{-25}^{+50}$           \ & 0.04          \ &  7.63        \\
$~~~~   ~~  0_2^+  $ &                          \ &                     \ & 5.4$_{-19}^{+21}$              \ & 0.14       \ &  0.00       \\
$~~~~   ~~  3_1^+  $ &                          \ &                     \ & 54$_{-20}^{+23}$              \ & \textbf{16.9}       \ &  \textbf{4.67} \\
$~~~~   ~~  2_3^+  $ &                          \ &                     \ & 46$_{-22}^{+31}$               \ & \textbf{44.5}       \ & \textbf{12.9}        \\
$0_5^+  ~~  2_1^+  $ &                           \ &                    \ &                           \ & 0.57         \ & 0.39    \\
$~~~~   ~~  2_2^+  $ &                           \ &                     \ &                          \ & 0.04       \ &   0.00    \\
$~~~~   ~~  2_3^+  $ &                           \ &                     \ &                          \ & \textbf{60.5}       \ &  \textbf{23.2}   \\
$4_4^+  ~~  4_1^+  $ &                           \ &                       \ & 0.43$_{-36}^{+88}$         \ & 0.07       \ &   0.39  \\
$~~~~   ~~  2_3^+   $ &                          \ &                       \ & 15$_{-11}^{+16}$           \ & 5.61       \ &       \\
$0_6^+  ~~  2_1^+  $ &                           \ &                    \ &                           \ & 1.82         \ &     \\
$~~~~   ~~  2_2^+  $ &                           \ &                     \ &                          \ & 3.32       \ &       \\
$~~~~   ~~  2_3^+  $ &                           \ &                     \ &                          \ & 5.14       \ &       \\
$2_{8}^+  ~~ 0_1^+   $ &                           \ &                      \ & 0.153$_{-53}^{+70}$           \ & 0.001     \ &         \\
$~~~~   ~~ 2_1^+   $ &                          \ &                     \ & 0.00003$_{-3}^{+434}$          \ & 0.00       \ &       \\
$~~~~   ~~  2_2^+  $ &                          \ &                     \ & 0.02$_{-2}^{+11}$           \ & 0.06          \ &           \\
$~~~~   ~~  2_3^+  $ &                          \ &                     \ & 0.08$_{-8}^{+61}$               \ & 0.04       \ &         \\
$~~~~   ~~  2_5^+  $ &                          \ &                     \ &                           \ & 12.5       \ &         \\
$8_1^+  ~~  6_1^+  $ & 105(23)              \ & 107$_{-26}^{+13}$            \ &                            \ & 57.2           \ &     \\
\hline
$\sigma      $     &               \ &            \ &                            \ & 11.2          \ & 15.5    \\
\hline
\hline
\end{tabular}}
\end{table}

Due to the shape coexistence \cite{Heyde11,Heyde16,Garrett22,Bonatsos232,Leoni24}, identification of the $0^{+}$ states of the normal states of the new spherical-like $\gamma$-soft nucleus is a critical step. Fig. 2(a) shows the evolutional behaviors of the lowest six $0^{+}$ states in $^{108-120}$Cd. In the experiments, the researchers have confirmed that two $0^{+}$ states in $^{108-120}$Cd (red and blue colour) do not belong to the normal states, but are the bandheads of the first and second intruder states respectively \cite{Garrett19,Garrett20} (please also see the experimental analysis in \cite{Wang25}). It is clear that the red and blue connected lines of the two $0^{+}$ states show the parabolic feature, a sign of the intruder state \cite{Heyde11,Heyde16,Garrett22,Bonatsos232,Leoni24}. $^{100,102}$Cd have more $0^{+}$ states, but they are not easily identified for the multi-shape coexistence \cite{Garrett19,Garrett20} and the weak coupling with the normal states \cite{Garrett12}. In $^{118,120}$Cd, the coupling interaction can be ignored \cite{Wang25}, but the higher $0^{+}$ states are absent. Therefore, in the Cd nuclei, the new spherical-like $\gamma$-soft spectra cannot be obtained explicitly.

Further analysis of the nearby nuclei is needed. The effects of the intruder states on the normal states can be ignored, and the experimental data should be more. I find that $^{106}$Pd satisfies these conditions. The boson number is $N=7$, which is the same as the ones in $^{110}$Cd and $^{118}$Cd. In Fig. 2(b), the lowest six $0^{+}$ states in $^{110}$Cd, $^{118}$Cd and $^{106}$Pd are shown. In $^{106}$Pd, the $0_{3}^{+}$ intruder state \cite{Peters16,Marchini22} is much higher than the $0_{2}^{+}$ state, which is like the case in $^{118}$Cd. In $^{118}$Cd, the value of $B(E2;0_{3}^{+}\rightarrow 2_{1}^{+})$ is $>$1.2 W.u., and in $^{106}$Pd, the value of $B(E2;0_{3}^{+}\rightarrow 2_{1}^{+})$ is around 2.41 W.u.. These results are very small. Thus the coupling interaction between the normal states and the intruder states can be also ignored in $^{106}$Pd. Importantly, compared with $^{118}$Cd, the $0_{4}^{+}$ state does not belong to the normal states too, which is not mentioned in previous studies.

If the normal states of $^{106}$Pd show the new spherical-like $\gamma$-soft spectra, there must be a $0^{+}$ state with energy that is twice the one of the $0_{2}^{+}$ state (two-times relationship). I indeed find this $0^{+}$ state, see the $0_{5}^{+}$ state of $^{106}$Pd in Fig. 2(b). The fit below will show that this is not a coincidence.

\section{Hamiltonian}

In the new model only the U(5) symmetry limit and the SU(3) symmetry limit are considered. Various $\gamma$-softness can emerge \cite{Wang22}. The Hamiltonian is as follows \cite{WangPt,Wang25}
\begin{eqnarray}
\hat{H}&=&c\{(1-\eta)\hat{n}_{d}+\eta[-\frac{\hat{C}_{2}[SU(3)]}{2N}+\alpha\frac{\hat{C}_{3}[SU(3)]}{2N^{2}}   \nonumber\\
&&+\beta\frac{\hat{C}_{2}^{2}[SU(3)]}{2N^{3}}+\gamma\frac{\Omega}{2N^{2}}+\delta\frac{\Lambda}{2N^{3}}]\},
\end{eqnarray}
where $c$, $\eta$, $\alpha$, $\beta$, $\gamma$, $\delta$ are fitting parameters. $\hat{n}_{d}=d^{\dag}\cdot \tilde{d}$ is the $d$ boson number operator. $\hat{Q}=[d^{\dag}\times\tilde{s}+s^{\dag}\times \tilde{d}]^{(2)}-\frac{\sqrt{7}}{2}[d^{\dag}\times \tilde{d}]^{(2)} $ is the SU(3) quadrupole operator and $\hat{L}=\sqrt{10}[d^{\dag}\times \tilde{d}]^{(1)}$ is the angular momentum operator. $\Omega$ is $[\hat{L}\times \hat{Q} \times \hat{L}]^{(0)}$ and $\Lambda$ is $[(\hat{L}\times \hat{Q})^{(1)} \times (\hat{L} \times \hat{Q})^{(1)}]^{(0)}$. $\hat{C}_{2}[SU(3)]$, $\hat{C}_{3}[SU(3)]$ are the SU(3) second- and third-order Casimir operators, respectively. In the SU(3) symmetry limit, the SU(3) second-order Casimir operator $-\hat{C}_{2}[\textrm{SU(3)}]$ can describe the prolate shape and the SU(3) third-order Casimir operator $\hat{C}_{3}[\textrm{SU(3)}]$ can present the oblate shape. The various rigid triaxial shapes can be obtained by the combinations of the square of the SU(3) second-order Casimir operator $\hat{C}_{2}^{2}[\textrm{SU(3)}]$ and $-\hat{C}_{2}[\textrm{SU(3)}]$, $\hat{C}_{3}[\textrm{SU(3)}]$.

\begin{table}[tbh]
\caption{\label{table:expee} The values of the quadrupole moments of some low-lying states in $^{106}$Pd in eb. $^{a}$From Ref. \cite{ensdf}, $^{b}$From Ref. \cite{Svensson95}, $^{c}$From Ref. \cite{Giannatiempo98,Giannatiempo18}.}
\setlength{\tabcolsep}{2.5mm}{
\begin{tabular}{ccccccc}
\hline
\hline
     &Exp. 1$^{a}$                     \ &Exp. 2$^{b}$                \ &Theo.                     \ &IBM-2$^{c}$   \ &    \\
 \hline
$Q_{2^{+}_{1}}    $ &$-0.51(7)$ \ & $-0.55(5)$  \ & $-0.47$  \ & $-0.42$\\
$Q_{4^{+}_{1}}    $ &         \ & $-0.77$$_{-8}^{+5}$   \ & $-0.97$  \ &$-0.60$\\
$Q_{6^{+}_{1}}    $ &         \ & $-1.11$$_{-10}^{+17}$   \ &$-1.43$ \ & $-0.70$ \\
$Q_{2^{+}_{2}}    $ &          \ & 0.39$_{-4}^{+5}$  \ & 0.31  \ & 0.28\\
$Q_{4^{+}_{2}}    $ &         \ & $-0.23$$_{-4}^{+14}$  \ & $-0.13$  \ & $-0.03$\\
$Q_{2^{+}_{3}}    $ &          \ & $-0.47$$_{-18}^{+7}$   \ & $-0.61$  \ & $-0.23$\\
\hline
$\sigma   $ &          \ &    \ & 0.175  \ & 0.232\\
\hline
\hline
\end{tabular}}
\end{table}

\begin{table}[tbh]
\caption{\label{table:expee} SU(3) decomposition of the $0_{1}^{+}$, $0_{2}^{+}$, $0_{3}^{+}$ and $0_{4}^{+}$ states in the new model, the U(5) symmetry limit and the O(6) symmetry limit.}
\setlength{\tabcolsep}{3.0mm}{
\begin{tabular}{ccccccc}
\hline
\hline
      new model             &  $0_{1}^{+}$                    \ & $0_{2}^{+}$               \ & $0_{3}^{+}$                  \ & $0_{4}^{+}$      \\
 \hline
(14,0)                      &0.036                        \ & 0.290                         \ & 0.478                        \ & 0.041           \\
(10,2)                     & 0.213                        \ & 0.349                           \ & 0.041                     \ &0.016             \\
(6,4)                         & 0.373                       \ & 0.009                         \ &0.107                        \ & 0.117              \\
(2,6)                       & 0.257                          \ & 0.301                          \ &0.261                          \ & 0.019            \\
(8,0)                       & 0.044                         \ & 0.006                         \ & 0.087                        \ & 0.138               \\
(4,2)                       & 0.059                        \ & 0.023                            \ & 0.024                         \ & 0.423             \\
(0,4)                    & 0.015                          \ & 0.019                             \ & 0.001                       \ & 0.150             \\
(2,0)                       & 0.003                        \ &0.003                           \ & 0.001                            \ & 0.096                   \\
\hline
\hline
      U(5)            &  $0_{1}^{+}$                    \ & $0_{2}^{+}$               \ & $0_{3}^{+}$                  \ & $0_{4}^{+}$      \\
 \hline
(14,0)                      &0.007                   \ & 0.116   \ & 0.112                         \ & 0.327                        \\
(10,2)                     &0.072                     \ &0.340  \ & 0.102                          \ & 0.017                   \\
(6,4)                         & 0.180                \ & 0.136      \ & 0.096                         \ &0.013                     \\
(2,6)                       &0.152                   \ & 0.003       \ & 0.485                          \ &0.054                        \\
(8,0)                       & 0.097                  \ & 0.019       \ & 0.086                        \ & 0.346                       \\
(4,2)                       & 0.304                   \ & 0.104     \ & 0.038                            \ & 0.052                        \\
(0,4)                    & 0.110                      \ & 0.116   \ & 0.001                             \ & 0.066                       \\
(2,0)                       & 0.078                    \ & 0.166   \ &0.080                           \ & 0.125                            \\
\hline
\hline
     O(6)            &  $0_{1}^{+}$                    \ & $0_{2}^{+}$               \ & $0_{3}^{+}$                  \ & $0_{4}^{+}$      \\
 \hline
(14,0)                      &0.469                       \ & 0.326                         \ & 0.122                        \ & 0.050           \\
(10,2)                     & 0.234                        \ & 0.004                           \ & 0.241                     \ &0.250             \\
(6,4)                         & 0.179                      \ & 0.203                       \ &0.056                        \ & 0.000              \\
(2,6)                       & 0.068                          \ & 0.186                         \ &0.053                          \ & 0.260            \\
(8,0)                       & 0.033                      \ & 0.170                        \ & 0.415                        \ & 0.013               \\
(4,2)                       & 0.011                        \ & 0.071                          \ & 0.049                         \ & 0.012             \\
(0,4)                    & 0.004                         \ & 0.028                            \ & 0.029                       \ & 0.075             \\
(2,0)                       & 0.002                        \ &0.012                         \ & 0.035                            \ & 0.340                  \\
\hline
\hline
\end{tabular}}
\end{table}

The two SU(3) Casimir operators $\hat{C}_{2}[SU(3)]$, $\hat{C}_{3}[SU(3)]$ have relationships with the quadrupole second or third-order interactions as follows:
\begin{equation}
	\hat{C}_{2}[\textrm{SU(3)}]=2\hat{Q}\cdot\hat{Q}+\frac{3}{4}\hat{L}\cdot\hat{L},
\end{equation}
\begin{equation}
	\hat{C}_{3}[\textrm{SU(3)}]=-\frac{4}{9}\sqrt{35}[\hat{Q}\times\hat{Q}\times\hat{Q}]^{(0)}-\frac{\sqrt{15}}{2}[\hat{L}\times\hat{Q}\times\hat{L}]^{(0)}.
\end{equation}
For a given SU(3) irreducible representation $(\lambda,\mu)$, the eigenvalues of the two Casimir operators under the group chain $\textrm{U(6)} \supset \textrm{SU(3)} \supset \textrm{O(3)}$ are expressed as
\begin{equation}
	<\hat{C}_{2}[\textrm{SU(3)}]>=\lambda^{2}+\mu^{2}+\mu\lambda+3\lambda+3\mu,
\end{equation}
\begin{equation}
	<\hat{C}_{3}[\textrm{SU(3)}]>=\frac{1}{9}(\lambda-\mu)(2\lambda+\mu+3)(\lambda+2\mu+3).
\end{equation}

The Hamiltonian (1) can be diagonalized using the SU(3) basis diagonalization Fortran code \cite{Wang08,Rosensteel90} with the $U(6)\supset SU(3) \supset SO(3)$ basis spanned by ${ | N(\lambda,\mu)\chi L \rangle}$, where $\chi$ is the branching multiplicity occurring in the reduction of $SU(3)\downarrow SO(3)$. The basis vectors are orthonormal, so the eigenstates can be expressed as
\begin{small}
\begin{equation}
\label{eq.2}
|N,L_{\zeta};\eta,\kappa,\xi\rangle
=\sum_{(\lambda,\mu)\chi}C^{L_{\zeta}}_{(\lambda,\mu)\chi}(\eta,\kappa,\xi)| N(\lambda,\mu)\chi L \rangle,
\end{equation}
\end{small}
where $\zeta$ is an additional quantum number distinguishing different eigenstates with the same angular momentum $L$ and $C^{L_{\zeta}}_{(\lambda,\mu)\chi}(\eta,\kappa,\xi)$ is the corresponding expansion coefficient.

The parameters of the new spherical-like $\gamma$-soft spectra in Fig. 1 are $\eta=0.5$, $\alpha=\frac{3N}{2N+3}$, $\beta=\gamma=\delta=0$, which is the simplest situation. To better fit the actual spectra characteristics in $^{106}$Pd, these parameters should have some small changes. In this paper, the parameters are $\eta=0.47$, $\alpha=\frac{3N}{2N+3}$, $\beta=0$, $\gamma=1.728$, $\delta=1.34$ and $c=1177.42$ keV. The $\hat{L}^{2}$ interaction should be added, and its coefficient $f$ is 37.04 keV. The $\eta$ is determined to allow the energies of the $0^{+}$ states to agree with the experimental data. The $\gamma$ and $\delta$ are to match the value of $B(E2;0_{2}^{+}\rightarrow 2_{1}^{+})$.

For understanding the B(E2) features, the B(E2) values are necessary. The $E2$ operator is defined as
\begin{equation}
\hat{T}(E2)=e\hat{Q},
\end{equation}
where $e$ is the boson effective charge. For the U(5) symmetry limit, the quadrupole operator is  $[d^{\dag}\times \tilde{d}]^{(2)}$, with which the quadrupole moment of the $2^{+}_{1}$ state in the U(5) limit is negative (not zero) \cite{Iachello87}. In the U(5) symmetry limit the quadrupole moment of the operator $\hat{Q}_{0}=[d^{\dag}\times\tilde{s}+s^{\dag}\times \tilde{d}]^{(2)}$ of the $2^{+}_{1}$ state is just zero. Thus, if $\hat{Q}=[d^{\dag}\times\tilde{s}+s^{\dag}\times \tilde{d}]^{(2)}-\frac{\sqrt{7}}{2}[d^{\dag}\times \tilde{d}]^{(2)}$ is used, in the SU(3) limit it is the $\hat{Q}$ while in the U(5) limit it plays the same role as the $[d^{\dag}\times \tilde{d}]^{(2)}$. Thus the $Q$ in the $E2$ operator is appropriate.

\section{Results for $^{106}$Pd}

\begin{figure}[tbh]
\includegraphics[scale=0.33]{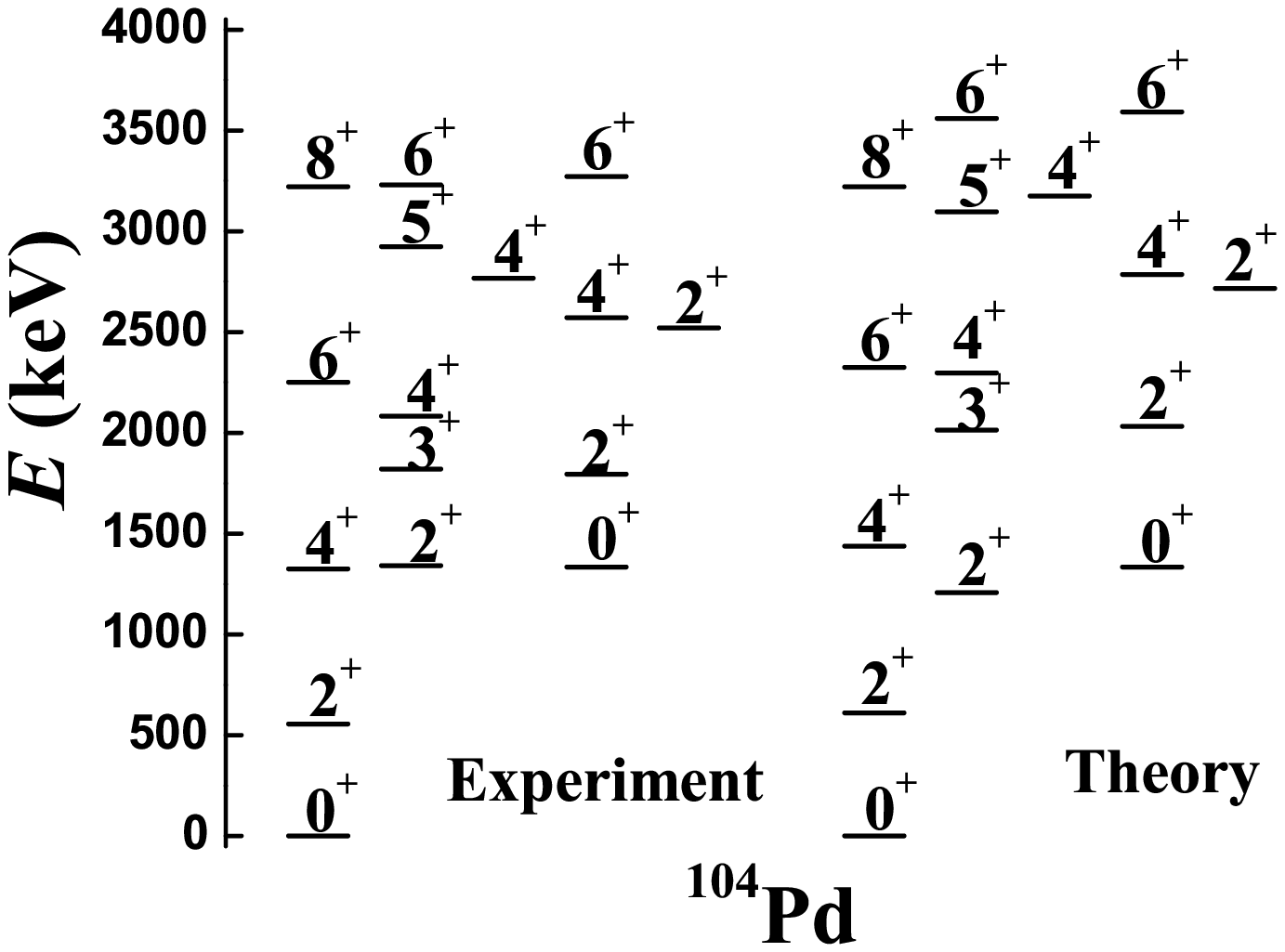}
\caption{The experimental and theoretical results of the low-lying levels of the normal states in $^{104}$Pd.}
\end{figure}

\begin{table}[tbh]
\caption{\label{table:expee}  Absolute $B(E2)$ values in W.u. for $E2$ transitions from the low-lying states in $^{104}$Pd and the value of the quadrupole moment of the $2_{1}^{+}$ state in $^{104}$Pd in eb. The theory has effective charge $e=2.082$ (W.u.)$^{1/2}$. $^{a}$From Ref. \cite{ensdf}, $^{b}$From Ref. \cite{Giannatiempo98,Giannatiempo18}.}
\setlength{\tabcolsep}{3.0mm}{
\begin{tabular}{ccccc}
\hline
\hline
$L_{i}  ~~~~L_{f}    $ &Exp.$^{a}$                                                  \ &Theo.   \ & IBM-2$^{b}$   \\
 \hline
$2_1^+  ~~~~ 0_1^+   $ & 36.9(19)                                         \ & 36.9    \ &  32.5  \\
$4_1^+  ~~~~  2_1^+  $ & 49(7)                                                 \ & 54.2       \ &  51.7   \\
$0_2^+  ~~~~2_1^+   $ & 13.2(13)                                          \ & 34.56       \ &     \\
$2_2^+  ~~~~ 2_1^+   $ & 21.8(17)                                           \ & 38.04       \ &  37.56  \\
$~~~~   ~~~~  0_1^+  $ & 1.29(10)                                   \ & 0.46       \ &  0.34   \\
$2_3^+  ~~~~ 0_1^+   $ & $>$0.066                                 \ & 0.023      \ &    \\
$4_2^+  ~~~~ 2_1^+   $ & 0.6(6)                            \ & 2.87   \ &  0.06  \\
$~~~~   ~~~~ 2_2^+   $ & 25(25)                                  \ & 21.59      \ &  26.3        \\
$~~~~   ~~~~  4_1^+  $ & 10(10)                                 \ & 11.68      \ &  19     \\
\hline
$Q_{2^{+}_{1}}    $ &$-0.47(10)$   \ & $-0.47$  \ & $-0.37$\\
\hline
\end{tabular}}
\end{table}

Fig. 3 shows the experimental and theoretical results of the low-lying levels of the normal states in $^{106}$Pd, up to the $10_{1}^{+}$ state and under 4000 keV. The degree of agreement between theory and experiment is striking. Most of the levels are almost uniformly located, except that the positions of the red levels are slightly higher. Experimentally, the three blue levels ($8^{+}$, $7^{+}$, $6^{+}$) have not been found yet, and are only marked according to the features of the level bands, but they can be believed to exist. Clearly, the third group levels $8_{1}^{+}$, $6_{2}^{+}$, $5_{1}^{+}$, $4_{3}^{+}$, $4_{4}^{+}$, $2_{4}^{+}$ and $0_{3}^{+}$ and the fourth group levels $10_{1}^{+}$, $8_{1}^{+}$, $7_{1}^{+}$, $6_{3}^{+}$, $5_{2}^{+}$, $6_{4}^{+}$, $4_{5}^{+}$, $3_{2}^{+}$, $2_{5}^{+}$ and $0_{4}^{+}$ really exist and can be easily found. (The order of labeling in the experiment agrees with Ref. \cite{Pd17}) The theoretical five lowest $0^{+}$ states all have the experimental correspondences.

Theoretically, the odd-even staggering effect in the $\gamma$ band can be seen, but experimentally, it is not prominent. This implies that the experimental levels with odd number angular moments $L$ are somewhat lower. In the two adjacent nuclei $^{104}$Pd and $^{108}$Pd, these experimental staggering effect can be clearly observed (see Fig. 4 and 5). I guess improving these fitting effects needs distinguishing between protons and neutrons of the new model. In \cite{Giannatiempo18}, these staggering effect can be better explained by the IBM-2.

Table I shows the absolute $B(E2)$ values for $E2$ transitions from the low-lying states in $^{106}$Pd. Qualitatively, the results of the theory and experiment fit at a good level, and there is no complete inconsistency. For the strong $E2$ transitions, they almost fit very well. The value of $B(E2;0_{2}^{+}\rightarrow 2_{2}^{+})$ (19(+7-3) W.u.) seems smaller than the theoretical result (84.6 W.u.), but in $^{108}$Pd, it's 47(+5-11) W.u. and in $^{112-116}$Cd, they are 99(16), 127(16) and even $3.0\times 10^{4}$(8) W.u. respectively. Overall, the new theory fits better than the calculation results in \cite{Giannatiempo98,Giannatiempo18} with the IBM-2. In the IBM-2, the value of $B(E2;2_{3}^{+}\rightarrow 0_{2}^{+})$ (12.3 W.u.) is much smaller than the experimental result (39(4) W.u.). In $^{108,110}$Pd, they are 35(+14-15) W.u. and 160(40) W.u. respectively. Clearly the values of $B(E2;4_{2}^{+}\rightarrow 3_{1}^{+})$, $B(E2;2_{5}^{+}\rightarrow 3_{1}^{+})$, $B(E2;2_{5}^{+}\rightarrow 2_{3}^{+})$ and $B(E2;0_{5}^{+}\rightarrow 2_{3}^{+})$ are all smaller. I look forward to more precise experiments with $^{106}$Pd. Table II shows the quadrupole moments of the $2_{1}^{+}$, $4_{1}^{+}$, $6_{1}^{+}$, $2_{2}^{+}$, $4_{2}^{+}$ and $2_{3}^{+}$ states. Clearly the new theory also agrees with the experimental result very well, and better than the calculation results in \cite{Giannatiempo98,Giannatiempo18}. From Table I, II, the deviation of the IBM-2 is about 1.36 times that of the new theory.

The IBM-2 calculation without higher-order interactions is insufficient to understand the systematic behaviors in the Cd nuclei \cite{Garrett08,Garrett12}, the B(E2) anomaly \cite{Garahn16} and the prolate-oblate shape asymmetric evolutions \cite{Wang23}. So it is insufficient for understanding the collective behaviour of atomic nuclei. In conclusion, the key conclusions of this paper can be obtained. This fit directly confirms the existence of the new spherical-like $\gamma$-soft spectra and denies the possibility of the existence of the spherical nucleus in Cd-Pd nuclei region.

\begin{figure}[tbh]
\includegraphics[scale=0.33]{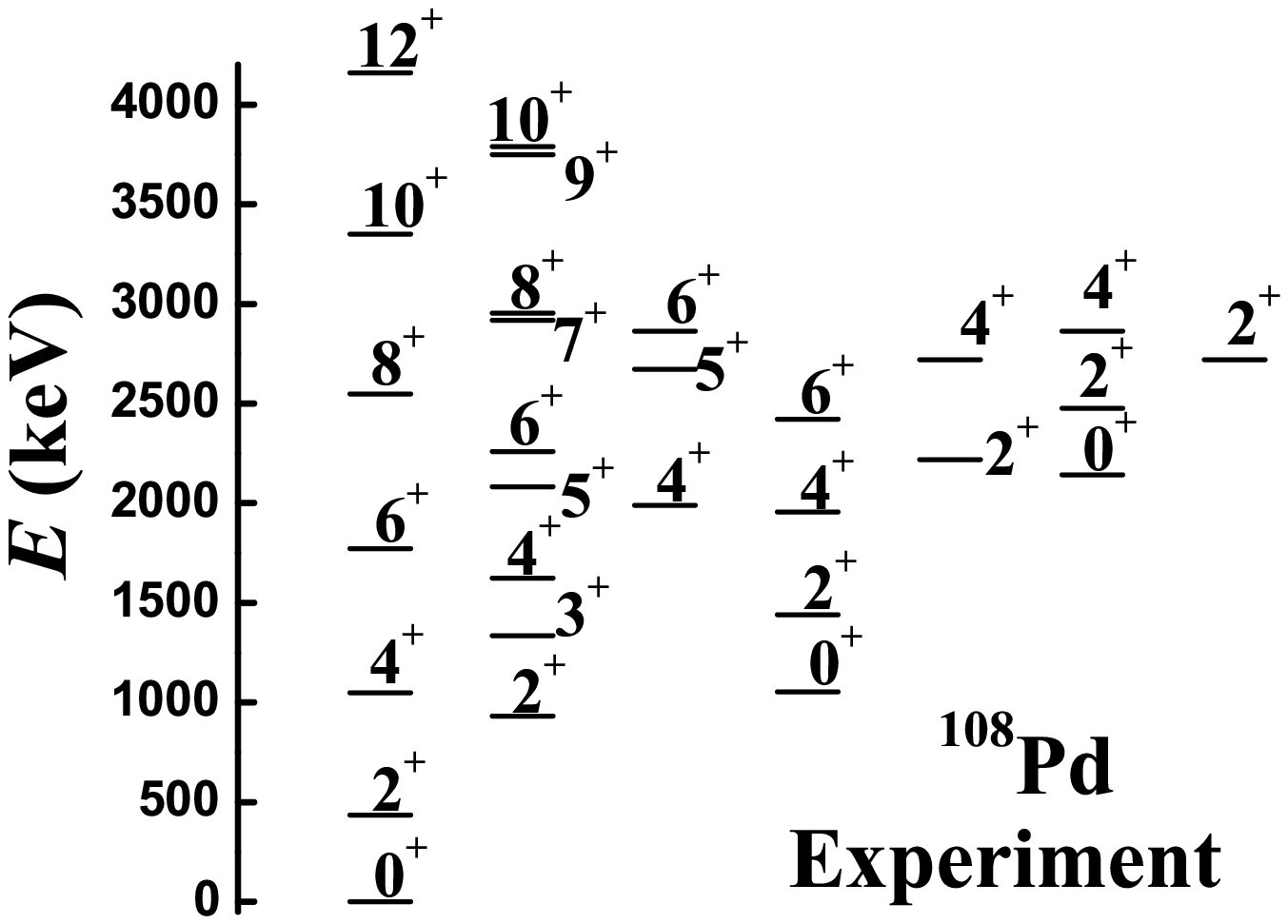}
\includegraphics[scale=0.33]{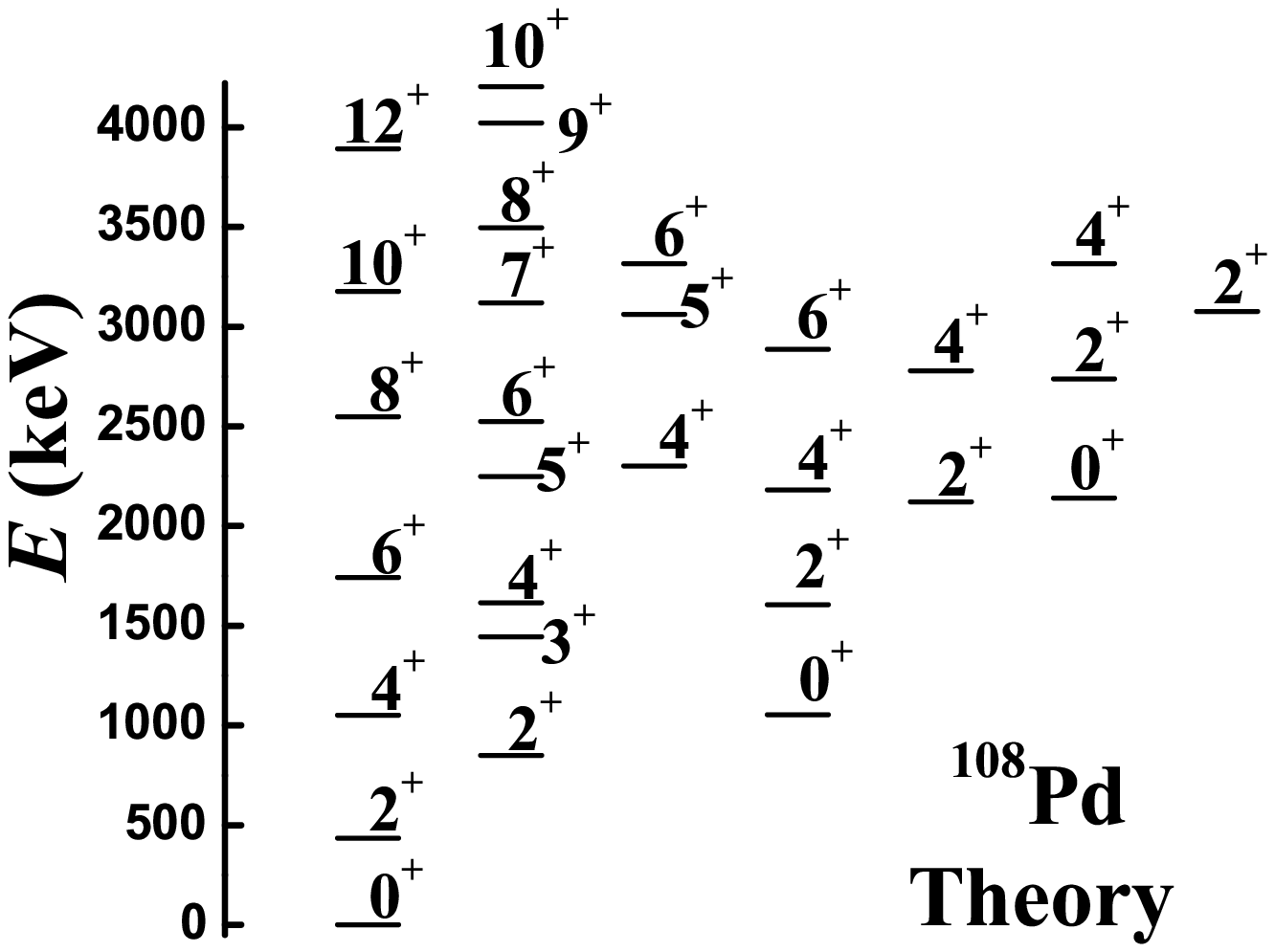}
\caption{The experimental and theoretical results of the low-lying levels of the normal states in $^{108}$Pd.}
\end{figure}

To further confirm that this is a new $\gamma$-soft mode, Table III presents the SU(3) decomposition of the $0_{1}^{+}$, $0_{2}^{+}$, $0_{3}^{+}$ and $0_{4}^{+}$ states in the new model, the U(5) symmetry limit and the O(6) symmetry limit. The intensity $|C^{L_{\zeta}}_{(\lambda,\mu)\chi}(\eta,\kappa,\xi)|^{2}$ of the SU(3) irrep. $(\lambda,\mu)$ are listed. Clearly, the decomposition components of the four $0^{+}$ states in the new model are very different from the ones in the U(5) symmetry limit and the O(6) symmetry limit. Taking the $0_{3}^{+}$ state as an example, the main components (larger than 0.1) in the new model are (14,0), (2,6) and (6,4) while they are (2,6), (14,0) and (10,2) in the U(5) symmetry limit and (8,0), (10,2) and (14,0) in the O(6) symmetry limit. For the $0_{1}^{+}$ state in the O(6) symmetry limit (O(6) $\gamma$-softness), the most important components are (14,0), (10,2), (6,4), satisfying $\lambda+2\mu=2N$. For the $0_{1}^{+}$ state in the new model, the main components are (10,2), (6,4) and (2,6), also satisfying the same condition.

\begin{table}[tbh]
\caption{\label{table:expee}  Absolute $B(E2)$ values in W.u. for $E2$ transitions from the low-lying states in $^{108}$Pd and the values of the quadrupole moments of some low-lying states in $^{108}$Pd in eb. The theory has effective charge $e=1.950$ (W.u.)$^{1/2}$. $^{a}$From Ref. \cite{ensdf}, $^{b}$From Ref. \cite{Giannatiempo98,Giannatiempo18}.}
\setlength{\tabcolsep}{3.0mm}{
\begin{tabular}{ccccccc}
\hline
\hline
$L_{i}  ~~L_{f}    $              \ &Exp.$^{a}$                      \ &Theo.   \ & IBM-2$^{b}$   \\
 \hline
$2_1^+  ~~ 0_1^+   $       \ &  50.4(15)                       \ &50.4    \ & 49.1  \\
$2_2^+  ~~ 2_1^+   $                 \ & 72(6)            \ & 62.6      \ &  51.8  \\
$~~~~   ~~  0_1^+  $                \ & 0.83(9)            \ & 0.12       \ &  0.93     \\
$0_2^+  ~~ 2_1^+   $         \ &   52(5)                     \ & 33.0      \ &     \\
$~~~~   ~~ 2_2^+   $          \ &    47$_{-11}^{+5}$                     \ &98.1       \ &      \\
$4_1^+  ~~  2_1^+  $                   \ &  76(9)                       \ & 74.1       \ &  79.5   \\
$~~~~   ~~ 2_2^+   $         \ &   1.21(14)                      \ & 7.15       \ &  0.54    \\
$2_3^+  ~~ 0_1^+   $          \ & 0.10$_{-4}^{+3}$           \ & 0.04      \ &  0.26       \\
$~~~~   ~~ 2_1^+   $  \ & 1.7$_{-2}^{+10}$           \ & 0.19      \ & 0.52     \\
$~~~~   ~~  2_2^+  $     \ & 11$_{-5}^{+4}$           \ & 5.86         \ &  2.47         \\
$~~~~   ~~  0_2^+  $             \ & 35$_{-15}^{+14}$         \ & 43.4      \ & 11.4      \\
$~~~~   ~~  4_1^+  $      \ & 6$_{-6}^{+7}$          \ & 22.4      \ &       \\
$4_2^+  ~~ 2_1^+   $           \ &   0.14(14)                     \ & 2.86   \ &  0.07  \\
$~~~~   ~~ 2_2^+   $          \ & 54(11)            \ & 33.9       \ &  43.4        \\
$~~~~   ~~ 2_3^+   $          \ & 3.6$_{-11}^{+27}$            \ & 9.24      \ &          \\
$~~~~   ~~  4_1^+  $            \ & 30(7)          \ & 23.7      \ & 28.4      \\
$4_3^+  ~~ 2_2^+   $           \ &   2.9(11)                    \ & 1.32   \ &   \\
$~~~~   ~~  4_1^+  $            \ & $<$1.8          \ & 0.14      \ &      \\
$~~~~   ~~  4_2^+  $            \ & 1.9$_{-11}^{+48}$           \ & 9.5     \ &      \\
$6_1^+  ~~ 4_1^+   $        \ & 107(13)           \ &91.9       \ &  93.5      \\
$8_1^+  ~~  6_1^+  $           \ &    148(17)                        \ & 102.2         \ &   93.9  \\
\hline
$Q_{2^{+}_{1}}$      \ &  $-0.58(4)$  \ & $-0.43$ \ & $-0.52$\\
$Q_{4^{+}_{1}}$       \ &  $-0.59(8)$   \ & $-0.97$ \ &$-0.70$\\
$Q_{6^{+}_{1}}$      \ &  $-0.53(13)$  \ &$-1.47$ \ & $-0.78$ \\
$Q_{2^{+}_{2}}$     \ & $0.55(6)$  \ & $0.26$  \ & $0.37$\\
$Q_{4^{+}_{2}}$      \ &  $-0.015$$_{-10}^{+6}$ \ & $-0.13$  \ & $0.007$\\
$Q_{2^{+}_{3}}$      \ &  $-0.41$$_{-18}^{+7}$   \ & $-0.58$  \ & $-0.25$  \\
$Q_{8^{+}_{1}}$     \ &  $-0.71$$_{-26}^{+13}$  \ &$-1.88$ \ & $-0.85$ \\
\hline
\end{tabular}}
\end{table}

To better understand the new spherical-like $\gamma$-softness in Pd nuclei, in this paper, $^{104}$Pd and $^{108}$Pd are also discussed. These studies further confirm this new $\gamma$-softness, but also offer some new problems. I perform lots of calculations around the parameters used for fitting $^{106}$Pd, and finally choose the same parameters $\eta=0.47$, $\alpha=\frac{3N}{2N+3}$, $\beta=0$, $\gamma=1.728$, $\delta=1.34$, and only the parameter $c$ and the coefficient $f$ of the $\hat{L}^{2}$ interaction are somewhat different.

\section{Results for $^{104}$Pd}

For $^{104}$Pd, $c=1317.46$ keV and $f=45.12$ keV. Fig. 4 presents the experimental and theoretical results of the low-lying levels of the normal states in $^{104}$Pd and Table IV shows the absolute $B(E2)$ values for $E2$ transitions from the low-lying states in $^{104}$Pd. The fitting effect still looks good. A key point different from $^{106}$Pd is that the experimental value of $B(E2;0_{3}^{+}\rightarrow 2_{1}^{+})$ in $^{104}$Pd is $>$25 W.u., much larger than the one 2.41 W.u. in $^{106}$Pd. Thus the coupling between the normal states and the intruder states in $^{104}$Pd can not be ignored, and a configuration-mixing calculation is required, which will be discussed in future. When the intruder states are considered, the theoretical value of $B(E2;0_{2}^{+}\rightarrow 2_{1}^{+})$ will be further reduced. In Ref. \cite{Droste22}, the B(E2) values of the levels in the yrast band of $^{104}$Pd were measured experimentally, and they show a peculiar behavior when the angular momentum $L$ increases, which is different from the ones in $^{106}$Pd.

\section{Results for $^{108}$Pd}

\begin{figure}[tbh]
\includegraphics[scale=0.33]{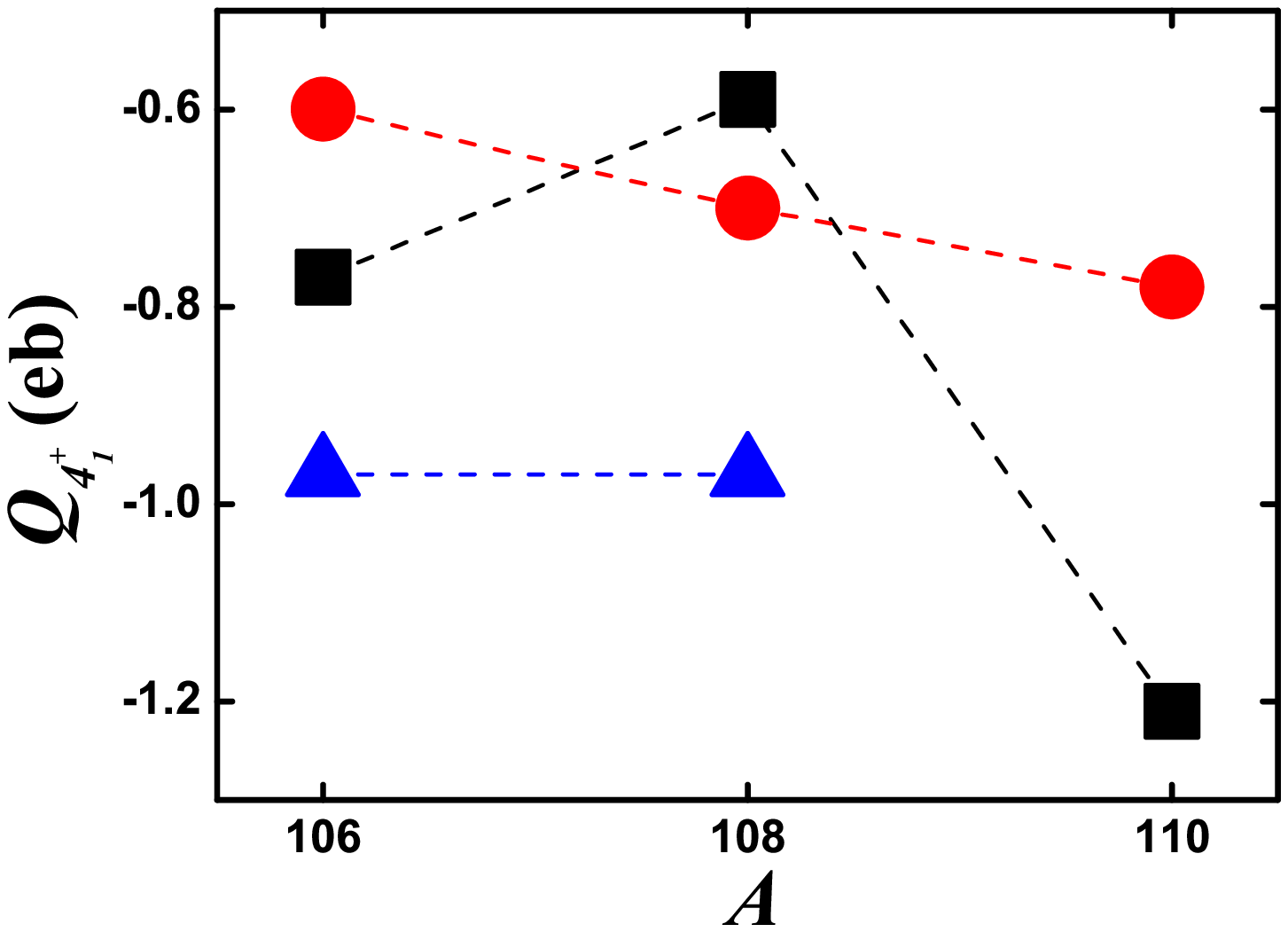}
\caption{The evolutional behaviors of the quadrupole moment $Q_{4_{1}^{+}}$ of the $4_{1}^{+}$ state in $^{106-110}$Pd. Black squares present the experimental data \cite{ensdf}, red circles presents the IBM-2 results \cite{Giannatiempo98,Giannatiempo18} and blue triangles presents the results of the new model.}
\end{figure}

For $^{108}$Pd, $c=1152.21$ keV and $f=27.40$ keV. Fig. 5 presents the experimental and theoretical results of the low-lying levels of the normal states in $^{108}$Pd. Overall, the fitting effect is very good, and the odd-even staggering effect in the $\gamma$ band can be shown. The characteristics of the new spherical-like $\gamma$-soft spectra are almost demonstrated. Clearly, the third group levels $8_{1}^{+}$, $6_{2}^{+}$, $5_{1}^{+}$, $4_{3}^{+}$, $4_{4}^{+}$, $2_{4}^{+}$ and $0_{3}^{+}$ really exists. Except for the $3_{2}^{+}$ and $0_{4}^{+}$ states, the $10_{1}^{+}$, $8_{1}^{+}$, $7_{1}^{+}$, $6_{3}^{+}$, $5_{2}^{+}$, $6_{4}^{+}$, $4_{5}^{+}$, $2_{5}^{+}$ states in the fourth group can be easily found. I look forward to doing more experimental researches on $^{108}$Pd and confirming the $3_{2}^{+}$ and $0_{4}^{+}$ states. Experimental odd-even staggering effect in $\gamma$-band is more prominent, so distinguishing between protons and neutrons is important.

Table V shows the absolute $B(E2)$ values for $E2$ transitions from the low-lying states in $^{108}$Pd. The fitting effect seems good. Insufficient experimental data prevent comparison with results from the IBM-2 \cite{Giannatiempo98,Giannatiempo18}. The quadruple moments of the low-lying states show that the deviations of theoretical results relative to the experimental ones are somewhat large. In the new model, this result is inevitable. In $^{108}$Pd, the values of the quadruple moments of the $4_{1}^{+}$ and $6_{1}^{+}$ states are $-0.59(8)$ eb and $-0.53(13)$ eb, larger than the ones $-0.77(+5-8)$ eb and $-1.02(+16-9)$ eb in $^{106}$Pd. The boson number of $^{108}$Pd is $N=8$. Thus from 106 to 108, the relative strength of the $4_{1}^{+}$ and $6_{1}^{+}$ states become smaller. In $^{110}$Pd, the values of $4_{1}^{+}$ and $6_{1}^{+}$ states are $-1.21(48)$ eb and $-0.98(39)$ eb, becoming more smaller than the ones in $^{108}$Pd. Thus from 108 to 110, the relative strength of the $4_{1}^{+}$ and $6_{1}^{+}$ states become large again. Fig. 6 presents the evolutional behaviors of the quadrupole moment $Q_{4_{1}^{+}}$ of the $4_{1}^{+}$ state in $^{106-110}$Pd. This is in fact an anomalous phenomenon (black squares). In the IBM-2 calculation \cite{Giannatiempo98,Giannatiempo18}, although the fitting effect is better, this anomaly is not presented (red circles).

In \cite{Zhao251}, the shape phase transition from the new spherical-like $\gamma$-soft mode to the prolate shape has been found and $^{108}$Pd was found to be the critical nucleus. Thus, $^{108}$Pd should be a softer nucleus, and the relative strength of the quadrupole moments of the $4_{1}^{+}$ and $6_{1}^{+}$ states should be smaller. Thus $^{108}$Pd is not the new spherical-like $\gamma$-soft nucleus, and only resembles it. A thorough  investigation to reproduce the features in $^{106-110}$Pd is vital for understanding the new model.

For the critical nucleus $^{108}$Pd, the experimental and theoretical evidence exists, but a quantitative coincidence is not yet available. This problem is very interesting and I guess this is a strong evidence for the need of distinguishing between protons and neutrons. In following studies, I will look for better results for fitting $^{106}$Pd and $^{108}$Pd in the entire parameter interval of the new model, and further distinguish between protons and neutrons, hoping to achieve a better fitting effect, and fully prove that $^{108}$Pd is the critical nucleus from the new spherical-like $\gamma$-soft to the prolate shape \cite{Zhao251}.

\section{Conclusion}

A conclusion can now be given that $^{106}$Pd is indeed the new spherical-like $\gamma$-soft nucleus. The levels of the states up to $10_{1}^{+}$ state under 4000 keV are all verified. The third and fourth group levels are all confirmed. The theoretical results of the B(E2) values and the quadrupole moments fit with the experimental data at a good level. $^{104,108}$Pd are also discussed, which further support the conclusions and provide new problems. A configuration-mixing calculation should be performed for $^{104}$Pd. A thorough discussion on the evolutions in $^{106-110}$Pd is also needed in future. To better include more energy levels, the new model distinguishing protons and neutrons will be considered. Combined with the fitting results of the new model to date, it can be believed that these investigations in the new model offer a more accurate description of the collective behaviors in atomic nuclei. One can be sure that the understanding of the shape evolutions in nuclear structure is undergoing a fundamental shift, just like the descriptions of Heyde and Wood: ``The emerging picture of nuclear shapes is that quadrupole deformation is fundamental to achieving a unified view of nuclear structure." and ``a shift in perspective is needed: sphericity is special case of deformation." \cite{Heyde16}.

\end{document}